\documentclass[
    ,final            
  ]
  {aipproc}
\layoutstyle{6x9}

\begin{document}

\title{Recent and Future Observations in the X-ray and Gamma-ray Bands: Chandra, Suzaku, GLAST, and NuSTAR}


\author{Greg Madejski}{
 address={Stanford Linear Accelerator Center and KIPAC, Stanford CA 94305, USA}
}

\begin{abstract}

This paper presents a brief overview of the 
accomplishments of the Chandra satellite that are shedding 
light on the origin of high energy particles in astrophysical 
sources, with the emphasis on clusters of galaxies. It also discusses 
the prospects for the new data 
to be collected with instruments recently launched - such as Suzaku - 
or those to be deployed in the near 
future, and this includes GLAST and NuSTAR.

\end{abstract}

\maketitle


\section{Introduction}

The last several years can be truly called the ``golden era'' of 
high energy astrophysics.  At the time of this meeting, Chandra 
and XMM-Newton were conducting imaging observations in the soft X-ray 
band;  RXTE was measuring the timing properties of variable celestial 
X-ray sources;  Integral covered the hard X-ray and soft gamma-ray 
regime;  and Swift, just launched, was effectively discovering and 
measuring properties of gamma-ray bursts.  All this resulted in 
tremendous advancements of our understanding of sources of high energy 
radiation - shedding light on the physical processes responsible for 
the particle acceleration in the Universe.  This paper focuses on selected 
results derived from Chandra data, but also presents the prospects 
for the near future - and this includes the recently launched Suzaku, 
and the approved satellite missions GLAST and NuSTAR.  

\section{Chandra X-ray Observatory}

Now entering the seventh year of its operation, Chandra X-ray Observatory 
was launched by the Space Shuttle into a deep, highly eccentric 
orbit which takes about 2 2/3 days to complete.  Chandra telescope is 
perhaps the most sophisticated X-ray optical system ever built, capable of 
imaging at the point-spread function better than $0.5''$.  
The focal plane instruments include the 
Advanced CCD Imaging Spectrometer (ACIS), an imaging instrument 
affording moderate-resolution ($\sim 100$ eV) spectral capability, 
with the combined mirror+detector 
effective area peaking at $\sim 700$ cm$^{2}$.  
The High Resolution Camera, a micro-channel plate detector, is capable of 
highest spatial resolution while still providing some spectral information.  
In addition, Chandra can perform high-resolution X-ray spectroscopy, 
via the use of low- as well as medium- and high-energy transmission gratings, 
read out by the Chandra imaging detectors.  The overall bandpass of the 
telescope spans $\sim 0.2 - 10$ keV.  

Over six years of operation, Chandra conducted a wide variety of 
scientific investigations.  Within our Galaxy, those range 
from studies of planets, star forming regions, 
normal stars, supernova remnants, binary accreting systems, 
globular clusters, as well as the Galactic Center.  Outside of our 
Galaxy, it also studied many 
normal and active galaxies, clusters of galaxies, and the content 
of the diffuse intergalactic medium, but the most important and 
unique study was that of the Cosmic X-ray 
Background (CXB).  Observations conducted with the Chandra 
and XMM-Newton satellites 
clearly indicated that the main contribution to the CXB in the 2 - 10 
keV band is indeed due to the superposition of unresolved active galactic 
nuclei (AGN), but at least the XMM data indicate that the situation is
less clear regarding the highest energies accessible to Chandra and XMM, 
above $\sim 5$ keV;  a more detailed discussion 
can be found in Worsley et al. (these proceedings).  
This is important, since the background spectrum, 
when plotted in the $E \times F(E)$ form, peaks somewhere near
30 keV.  With this, it will be important to conduct sensitive observations 
in the hard X-ray band, and the currently planned hard X-ray imaging satellite 
NuSTAR has this as one of its main goals.  

\begin{figure}
  \includegraphics[height=.25\textheight]{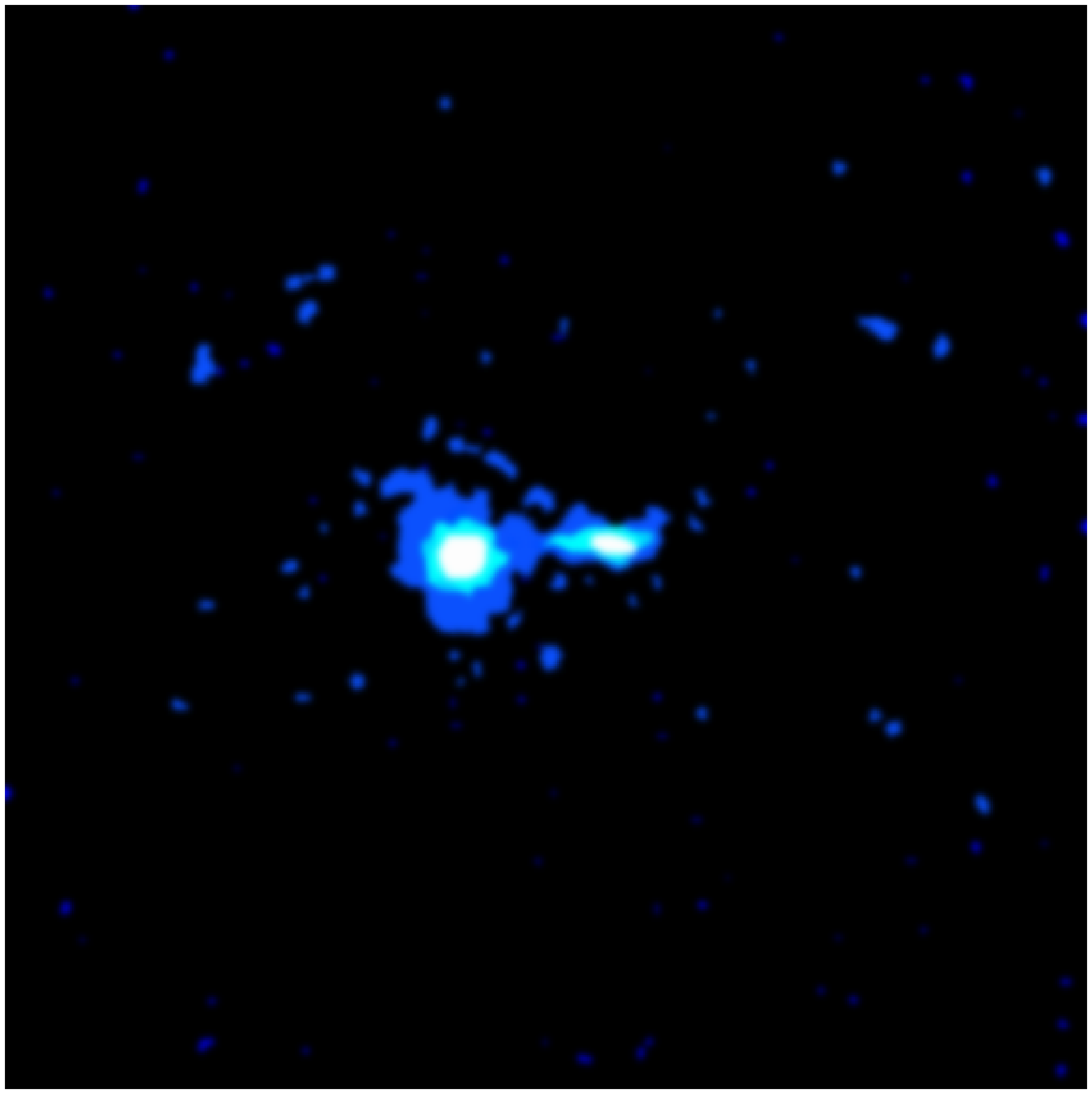}
\includegraphics[height=.25\textheight]{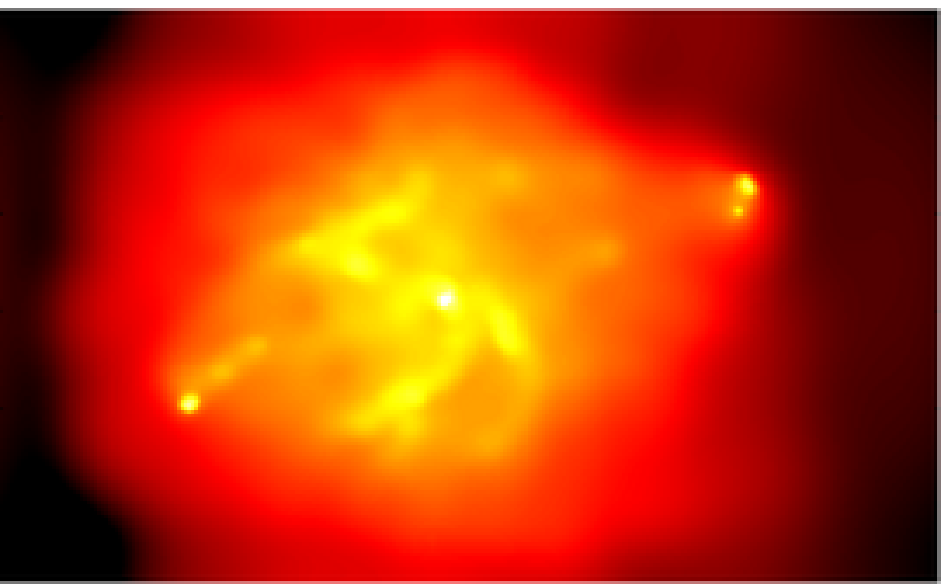}
  \caption{Left:  X-ray image of the quasar PKS 0637-752, revealing the 
kpc-scale X-ray emitting jet (data published in Schwartz et al. 2000;  
figure credit:  Chandra X-ray Center and NASA).  
Right:  Chandra X-ray image of the radio galaxy Cyg A, revealing 
X-ray jets (Figure source:  Chandra X-ray Center;  figure credit: 
NASA;  data published by Wilson, Young, and Shopbell 2000)} 

\end{figure}

Chandra's accomplishments regarding various astrophysical sources, 
especially neutron stars, supernova remnants, and active 
galaxies, were discussed extensively at the Torun meeting
and a number of results are discussed in various papers contained in 
these proceedings.  In particular, well-known is the discovery of 
large-scale (kpc) X-ray emitting jets, especially in the highly 
luminous, distant quasars (see Fig. 1);  this is discussed more
extensively by Stawarz (these proceedings).  There, the best scenario for 
production of X-ray radiation is Compton upscattering of the Cosmic 
Microwave Background radiation by energetic, relativistic electrons in 
the jet, but regardless of the radiation mechanism, 
the radiating particles must be accelerated locally.  The data for extended 
AGN jets are likely to hold many clues as to the processes responsible for 
formation of these jets, their composition, 
as well as the transport of energy from the 
central source -- presumably an accreting supermassive black hole -- 
to the distances comparable (or larger) to the size of the host galaxy;  
this subject is discussed in Sikora et al. (these proceedings).  
Below I focus on one area not covered by other 
contributions, namely clusters of galaxies.  

\subsection{Clusters of galaxies observed with Chandra and XMM}

The superb quality of the Chandra mirrors produced exquisite images
of extended celestial X-ray sources, and this is certainly true for 
clusters of galaxies.  The images revealed rich structure, associated with 
the process of formation and evolution 
of clusters - such as merging activity, or 
interaction of the central AGN with the intra-cluster medium.  The 
availability of high-quality strong and weak lensing data allows 
detailed comparison of the gravitating mass against the properties of the 
X-ray emitting gas.  In general, the total (dark + baryonic matter) masses 
inferred from weak lensing agree with those inferred from X-ray observations, 
but discrepancies betwen X-ray and strong lensing masses are 
commonly inferred in non-relaxed systems, 
which are presumably still forming, and are not in dynamical equilibrium.  
Particularly compelling results were inferred from the Chandra observations of 
the ``bullet cluster'' (1E0657-56;  Fig. 2) by Markevitch et al. (2004) 
and Clowe et al. (2004).  Those authors report that the cluster 
is undergoing a high-velocity $\sim 4500$ km s$^{-1}$ merger, 
evident from the spatial distribution of the 
hot, X-ray emitting gas, but this gas lags behind the 
subcluster galaxies.  Furthermore, the dark matter clump, revealed 
by the weak-lensing map, is coincident with the collisionless galaxies, but 
lies ahead of the collisional gas.  This -- and other similar observations --
allow good (and interesting) limits on the cross-section of the 
self-interaction of dark matter.  Observations such as this - 
and many other, similar ones - are bound to reveal the details of 
the process of cluster formation, and in particular, of the heating
of the intra-cluster gas.  

\begin{figure}
\includegraphics[height=.45\textheight]{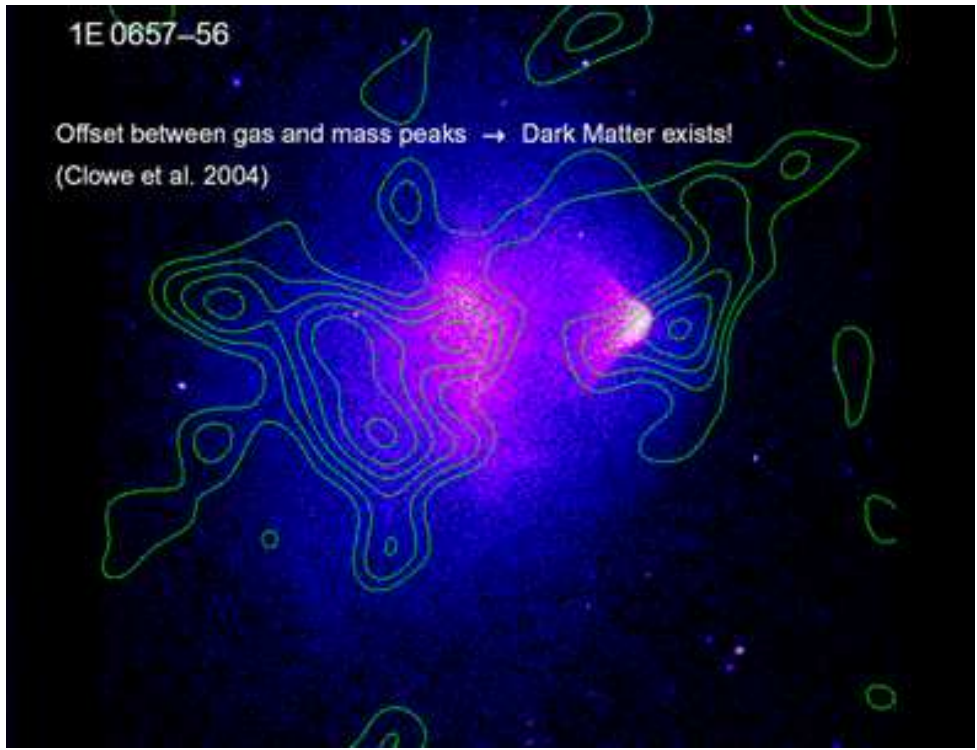}
\caption{Chandra X-ray image of the ``bullet cluster'' 
(1E0657-56) overlaid on the 
inferred distribution of the matter (drawn as green contours) -- 
presumably associated with the cluster -- that is likely 
responsible for lensing of light from background galaxies 
(figure courtesy of Maxim Markevitch, priv. comm., with permission).  
Data are from Markevitch et al. 2004 and Clowe et al. 2004.  
The offset between the X-ray - emitting 
matter and lensing matter distributions provides important 
evidence for existence of dark matter.  }

\end{figure}

Another important aspect of the cluster evolution is the 
interaction of the central AGN, present in many clusters 
of galaxies, with the intergalactic gas.  Particularly 
impressive are the Chandra images of the Perseus cluster, 
revealing a cavity as well as ripples, presumably 
due to the interaction of the outflow (jet?) from the active galaxy 
NGC 1275 with the ICM of the cluster (Fabian et al. 2003;  see Fig. 3).  
Such interaction might be responsible for the absence of cool cores 
in the cluster X-ray data;  those in turn were expected 
on theoretical grounds, since the characteristic cooling time 
of the cluster gas should be relatively rapid, certainly much 
shorter than Hubble time.  Instead, the high spectral resolution 
XMM-Newton data revealed that the clusters are generally devoid of 
cold cores, but instead, those cores have temperatures that are 
typically $\sim 1/3$ of the cluster average 
(Peterson et al. 2003).  

\begin{figure}
\includegraphics[height=.35\textheight]{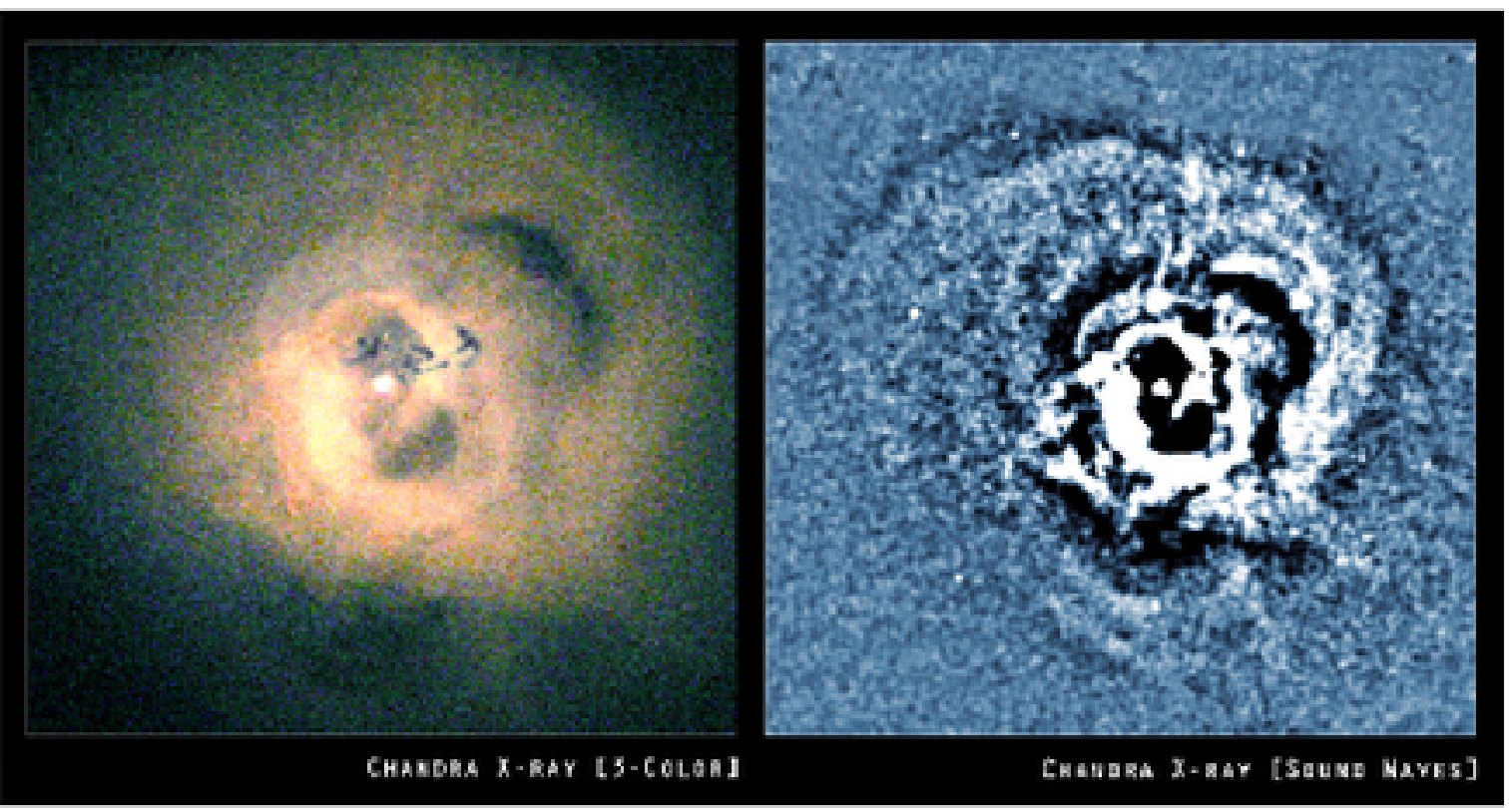}
\caption{Deep Chandra X-ray image of the Perseus cluster 
(figure source:  Chandra Science Center;  figure credit:  
NASA;  data published in 
Fabian et al. 2003).  The left panel shows the smoothed image, 
while the right panel shows the image in high contrast.  The 
structure seen in both images is real, presumably due to the interaction
of the central AGN (NGC 1275) with the cluster gas.}

\end{figure}

Chandra continues to conduct observations of clusters, 
and the goal is two-fold.  On one hand, the spatially-resolved spectroscopy 
allows a good insight into the formation of clusters and their evolution.  
Besides answering the obvious question:  ``what are the processes that 
heat the intra-cluster gas?'' - such data should reveal some clues 
as to the origin of 
non-thermal population of energetic (relativistic) particles, inferred 
from radio observations.  Radio emission is non-thermal, presumably due to 
synchrotron emission by relativistic particles in the intra-cluster 
magnetic field.  Such particles might be accelerated in shocks, 
resulting from galaxy collisions during the formation of the cluster - and 
this is probably what the data for the ``bullet cluster'' (Fig. 2) reveal.  
If such energetic, non-thermal particles are indeed present, 
they should Compton-upscatter the 
Cosmic Microwave Background to hard X-ray energies.  This should be detectable 
as a departure form thermal emission, and is best studied in the 
hard X-ray band, with instruments such as NuSTAR, described in 
subsequent sections.  

Armed with good understanding of cluster evolution, we can use them as 
cosmological tools, since clusters should provide a ``fair sample'' of 
the matter content of 
the Universe.  This very well complements the Cosmic Microwave Background 
and supernova data, providing somewhat orthogonal constraints towards 
the determination of cosmological parameters.  
There are several fruitful avenues to use the 
cluster data in cosmology.  One includes measurements of absolute distances 
to clusters, providing the value of Hubble constant.  This has been 
successfully attempted by a variety of workers, using Sunyaev-Zeldovich 
effect (see, e.g., Reese et al. 2004) or X-ray - inferred temperatures 
as well as masses of both dark an luminous matter 
(see, e.g., Allen et al. 2004).  The other relies on 
determining the evolution of cluster density as a function of cosmological 
time, and since this is related to the growth of structure in general, 
it depends sensitively on cosmological parameters.  In both cases we need 
to measure the cluster masses, which is relatively straightforward 
if a cluster is dynamically relaxed but more challenging 
for the more complex, merging systems:  this requires a good 
knowledge of the processes that lead to formation of the cluster.  

\section{Future High-Energy Astrophysics Missions}

While the current data are providing us with spectacular 
high-energy observations of celestial sources, many questions remain. 
Perhaps the most pressing is the dynamical state of the X-ray emitting gas 
in a wide range of celestial sources.  
This is best studied with high resolution X-ray spectroscopy, where the 
Doppler shifts of spectral features provide information about the 
kinematics of the gas.  Another somewhat under-explored area regards 
the details of the hard X-ray and soft gamma-ray emission, bound to reveal 
the nature of the fundamental processes converting gravitational energy of 
material being accreted onto a compact object into radiation.  This must be 
associated with heating and acceleration of particles to relativistic 
energies, and is best probed at energies beyond where the thermal 
processes dominate, which for many sources is around 10 keV.  Finally, 
the very successful EGRET instrument, operational in the 1990s on-board of the 
Compton Gamma-ray Observatory, revealed many different classes of 
energetic ($\sim$ GeV) gamma-ray emitters, but better understanding 
of those sources requires data with more sensitive instruments.  The future 
appears promising:  Suzaku should usher sensitive, high-resolution X-ray 
spectroscopy, with energy resolution at least 10 times better than the 
current ``workhorses'' of X-ray astronomy, CCDs;  NuSTAR will, for 
the first time, provide us with X-ray optics for imaging data for celestial 
sources in the 10 - 80 keV band, improving the ultimate sensitivity by a 
factor of at least several hundred;  and GLAST will have significantly 
improved sensitivity and broader bandpass than its predecessor EGRET.  
The capabilities of those three new missions are briefly discussed below.  

\subsection{Suzaku (Astro-E2)}

This satellite is the next in a series of Japanese X-ray astronomy missions, 
following the successful Hakucho, Tenma, Ginga, and Asca, and will be launched 
about 2 weeks after the conclusion of the Torun conference.  Developed in 
collaboration with US institutions, it features three types of 
instruments.  The premier instrument, the X-ray Spectrometer (XRS), 
is a non-dispersive, cryogenically cooled detector, operating at 
$\sim 60$ mK, with the expected cryogen lifetime of $\sim 2.5$ years.  
The detector, developed by the Goddard Space Flight Center 
consists of 32 independent sensors (pixels), where 
the interaction of an X-ray with the sensor results in a temporary 
rise of the sensor temperature roughly proportional to the energy 
of the incoming X-ray.  The XRS is located in the focal plane of an 
X-ray mirror with a modest (1 arc min) point-spread function.  
The mirror / detector combination provides sensitivity over the 
0.3 - 10 keV bandpass with effective area $>$ 100 cm$^{2}$ over 
the 0.8 - 8 keV band, providing an energy resolution better than 
10 eV.  The primary objective of this instrument is to study extended 
(rather than point-like) sources such as supernova remnants 
and clusters of galaxies, which cannot be reliably studied with 
diffraction gratings such as those onboard XMM-Newton or 
Chandra.\footnote{Unfortunately, as these proceedings were undergoing the 
editorial process, the cryogenic system, responsible for cooling the XRS, 
failed and resulted in the loss of the cryogen.  }

\begin{figure}
\includegraphics[height=.5\textheight]{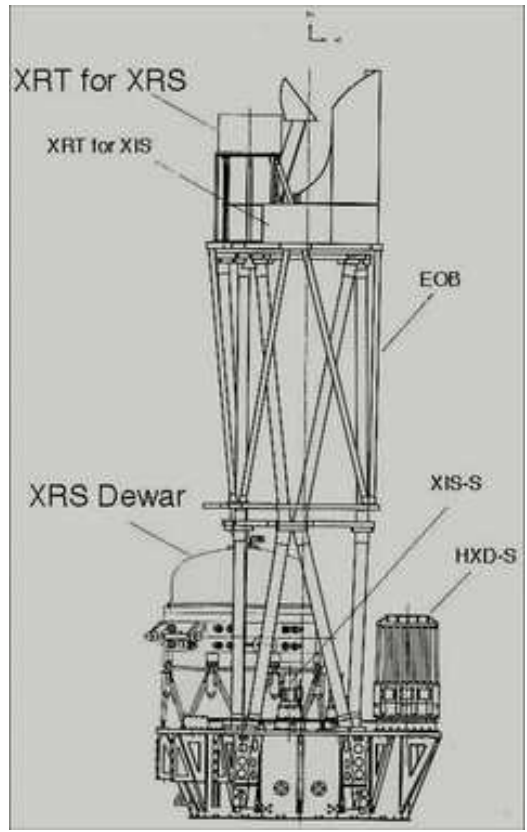}
\caption{Schematic illustrations of the Suzaku satellite, 
indicating its instruments:  the X-ray Spectrometer (XRS), the 
X-ray Imaging Spectrometer (XIS), and the Hard X-ray 
Detector (HXD).  XRT refers to the X-ray Telescope, and EOB to the Extendable 
Optical Bench.  Figure credit:  JAXA/ISAS}

\end{figure}

Besides the XRS, Suzaku features four more X-ray mirrors, each with 
the PSF of $\sim 1$ arc min, similar to the mirror for the XRS.  
Each mirror focuses X-rays onto an X-ray Imaging Spectrometer (XIS) 
featuring a CCD chip (similar in performance to those flown on the 
Chandra satellite), covering $18 \times 18$ arc min, and featuring the 
effective area of $\sim 400$ cm$^{2}$ at 1.5 keV.  The purpose 
of the XISs is to provide large effective area for detailed studies 
of temporally-resolved spectra, and will be particularly valuable after the 
depletion of the XRS cryogen.  This capability, coupled with the 
relatively large solid angle subtended by the CCDs, will be useful for 
mapping of substantial regions of the sky, or for monitoring of flux and
spectral variability of X-ray sources.  

The third kind of instrument on-board Suzaku is the Hard X-ray Detector, 
a non-imaging, collimated device with effective area of nearly 200 cm$^{2}$, 
sensitive over the $\sim 8 - 700$ keV range.  It actually consists 
of two separate sets of sensors:  the energy range $\sim 8 - 50$ keV 
is covered by PIN silicon diodes (with the field of view of 
$1/2 \times 1/2$ degree), while over $50 - 700$ keV, the sensors 
are GSO scintillators (with the field of view of $\sim 
4 \times 4$ degrees), physically located behind the PIN diode, 
which becomes transparent to X-rays with $E > 50$ keV.  The detector 
features a well-type design, meaning that the collimator (as well as the 
back and the sides of the instrument) are constructed of anti-coincidence 
shield, dramatically reducing the particle-induced background.  

More detailed description of Suzaku instruments 
as well as full references can be found at 
http://heasarc.gsfc.nasa.gov/docs/astroe/

\subsection{GLAST}

Gamma-Ray Large Area Space Telescope (GLAST) is under construction for 
a launch in 2007.  Its goal is to follow the successful energetic 
gamma-ray detector EGRET flown on-board of the Compton Gamma-Ray 
Observatory towards a study of celestial sources of gamma-ray 
radiation in the 30 MeV - 300 GeV range.  Such a broad range, 
especially at the high energy end, should allow some overlap of 
sensitivity with the ground-based Cherenkov telescope arrays, a 
feature highly desirable for cross-calibration.  The main instrument, 
the Large Area Telescope (LAT), is essentially a very sophisticated 
particle tracker, and is currently undergoing integration and test at 
the Stanford Linear Accelerator Center.  It determines the direction 
and energy of an incident gamma-ray by tracking the products of the 
interaction of the gamma-ray with converter material; the particle 
tracking uses silicon strip detectors.  The tremendous increase in 
sensitivity (as compared to EGRET) is mainly due to three separate 
improvements.  One is the larger effective area of the instrument, 
which, after accounting for instrumental inefficiencies should be at 
least 8000 cm$^{2}$ at 1 GeV.  Another is the fact that the detector 
has wide angle of acceptance, simultaneously ``seeing'' about 2  
steradians of the sky.  Finally, the third improvement is the 
significantly better ability to localize the direction of the incoming 
gamma-rays, which in turn allows for fewer background events or truly 
diffuse gamma-rays in the region corresponding to the PSF.  The 
current design goals call for the on-axis Point-Spread Function of 
$< 3.5$ degrees at 100 MeV and $< 0.15$ degrees at 10 GeV (localization 
of celestial sources should be considerably better than these values, 
via centroiding).

\begin{figure}

\includegraphics[height=.4\textheight]{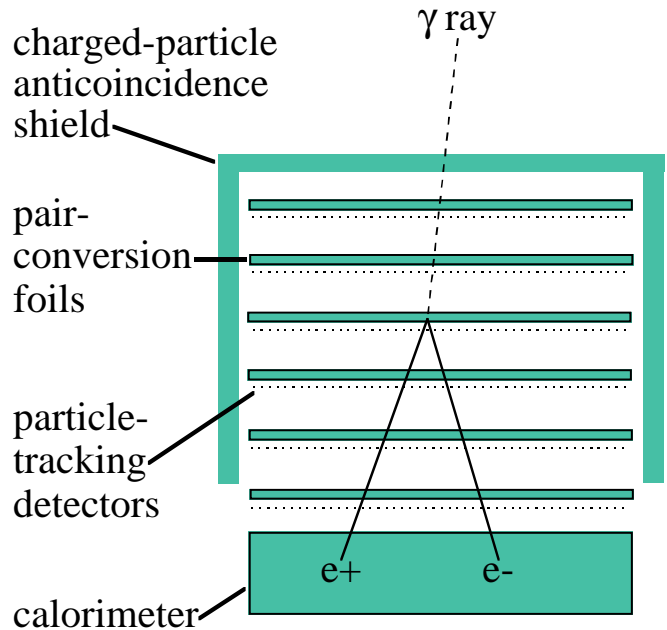}
\caption{Schematic illustration of the principle of operation of the 
Large Area Telescope, the main instrument onboard of the GLAST satellite.  
The calorimeter is capable of determining the total energy 
of particles resulting from the interaction of the gamma-ray 
with the converter, and should allow for energy resolution of the 
instrument of $\sim 10$\%.  Figure credit:  NASA and SLAC}

\end{figure}

The LAT detector will be always pointing away from the Earth, 
thus maximizing the exposure to the celestial gamma-rays.  With the large 
solid angle of acceptance of 2 steradian, the GLAST LAT is particularly 
well-suited to monitoring of gamma-ray flux from celestial sources, since 
every patch of the sky will be in the field of view of the detector for 
at least a fraction of each day.  This will be particularly useful 
for studies of variable gamma-ray sources such as the jet-dominated 
active galactic nuclei (blazars) that proved to be the most numerous 
point-like sources in the EGRET data.  However, since any better understanding 
of those sources will require observations at many different 
spectral bands, the real challenge will be to obtain adequate radio, IR, 
optical, and X-ray monitoring for a good number of sources, since 
all those bands generally 
allow studies of celestial sources ``one at a time.''  
Still, at least in the X-ray band, the Suzaku above, and NuSTAR below, 
are very well-suited to this task.  

\subsection{NuSTAR}

While the improvement in sensitivity of astrophysical 
instruments operating at energies below 10 keV has been remarkable, mainly 
via the Chandra and XMM-Newton observatories, above 10 keV, 
recent progress was only modest.  This is mainly due to 
experimental limitations:  the in-orbit 
background (both cosmic X-ray, and particle-induced) is considerable, 
and thus increasing the area of a {\sl detector} for simple, non-imaging, 
collimated instruments, is only of limited use.  
To achieve a large sensitivity gain, it is necessary to image the 
sky, preferably via the use of X-ray focusing optics.  This approach 
is of course the reason for the dramatic 
improvements below 10 keV:  it allows a 
reduction of the size of the detector (thus reducing background) 
but still, allows the collecting area -- that of the telescope -- to be only 
limited by the cost.  

\begin{figure}

\includegraphics[height=.32\textheight]{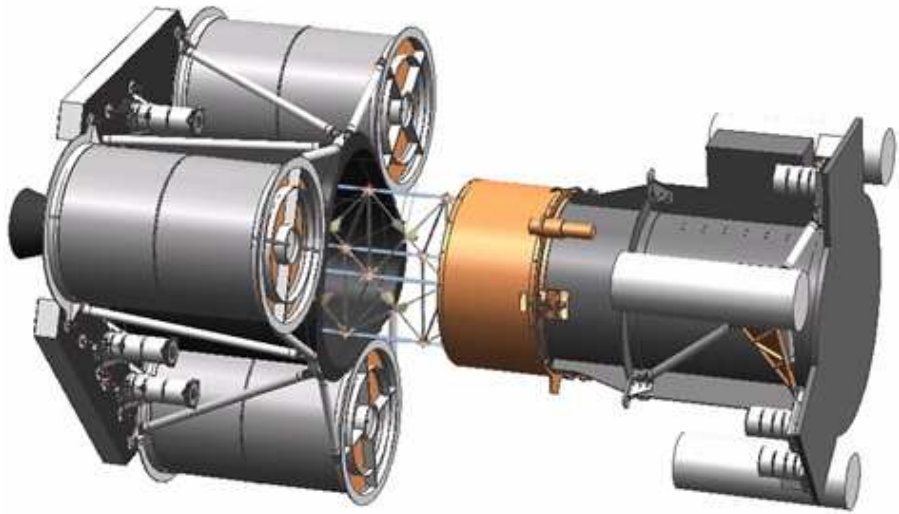}
\caption{Schematic illustrations of the NuSTAR satellite, 
designed to image the celestial X-ray sources in the hard X-ray band.  
The payload will extend in orbit, allowing for 10 meter separation 
between the X-ray mirrors and detectors.  Figure credit:  F. Harrison, 
NuSTAR PI, and NASA/JPL}

\end{figure}

Above 10 keV, the recent major breakthrough is the 
application of multi-coated layers to grazing incidence X-ray optics, and
this technique allows for focusing of hard X-rays.  Success of this
approach has been demonstrated by the recent flight of the Caltech - 
Livermore - Columbia - DSRI 
balloon payload HEFT (Harrison et al. 2000).  This payload included 
a mirror capable of focusing hard X-rays onto a CdZnTe pixellated detector, 
with energy resolution of $\sim 1$ keV.  
A similar approach is planned for the satellite-based instrument NuSTAR, 
also developed by the HEFT team but also including  
Stanford, UCSC, 
and JPL:  it will feature three multi-coated grazing incidence 
telescopes capable of focusing X-rays 
with energies 10 up to 80 keV onto detectors similar to those flown on HEFT.  
Since the grazing incidence angles required for reflection at the
hard X-ray  energies are shallow, a long focal length is required for an 
appreciable collecting area of the mirrors.  To this end, NuSTAR will
feature an extendible optical bench, allowing for a 10 m focal length, 
even though the total payload in the launch configuration will be no more  
than $\sim 1 \times 1 \times 1$ meter.  
The anticipated total effective area of all three telescope/detector 
systems is at least 300 cm$^{2}$ at 30 keV, with the point-spread function of 
$40''$.  This will result in a dramatic reduction of background, and 
will improve the sensitivity by a factor of many hundreds as compared 
with the previously flown, non-imaging hard X-ray instruments.  

NuSTAR will fly in 2009 as a part of NASA's Small Explorer program.  
It has three main science goals:  first, to take the census of black holes 
at all sizes.  This includes those powering active galaxies that 
contribute to the Cosmic X-ray background - and hopefully will result 
in resolving the questions brought by the current $E < 10$ keV 
data (see, e.g., Worsley et al., these proceedings).  Another goal
is to study the process of elemental enrichment of the Universe, 
by measuring spectra of young supernova remnants in a search for $^{44}$Ti 
emission lines:  the content and kinematics of this 
element in the ejecta has important implication on the details 
of supernova explosion.  Finally, NuSTAR will work closely with 
GLAST, providing the unique capability of monitoring the hard X-ray flux 
from the gamma-ray emitting variable celestial sources such as the 
jet-dominated active galaxies.  This is only a short list of 
scientific goals:  others include search for, and mapping the 
continuum hard X-ray emission in clusters of galaxies (revealing the details 
of non-thermal particle population), supernova remnants (in search for the 
origin of most energetic particles accelerated in the SNR and related 
to the Galactic cosmic rays).  

In summary, the ``golden era'' of high energy astrophysics, 
ushered by great instruments and resulting in spectacular discoveries, 
is likely to continue for many years to come.



\IfFileExists{\jobname.bbl}{}
 {\typeout{}
  \typeout{******************************************}
  \typeout{** Please run "bibtex \jobname" to optain}
  \typeout{** the bibliography and then re-run LaTeX}
  \typeout{** twice to fix the references!}
  \typeout{******************************************}
  \typeout{}
 }



\end{document}